\documentclass[conference]{IEEEtran}
\IEEEoverridecommandlockouts
\usepackage{cite}
\usepackage{amsmath,amssymb,amsfonts}
\usepackage{algorithmic}
\usepackage{graphicx}
\usepackage{textcomp}
\usepackage{xcolor}
\usepackage{multirow}
\def\BibTeX{{\rm B\kern-.05em{\sc i\kern-.025em b}\kern-.08em
    T\kern-.1667em\lower.7ex\hbox{E}\kern-.125emX}}
\begin{document}

\title{PAENet: A Progressive Attention-Enhanced Network for 3D to 2D Retinal Vessel Segmentation\\
\thanks{$\dagger$ Corresponding authors.}
}
\author{Zhuojie Wu\textsuperscript{1}, Zijian Wang\textsuperscript{1}, Wenxuan Zou\textsuperscript{1}, Fan Ji\textsuperscript{2}, Hao Dang\textsuperscript{3}, Wanting Zhou\textsuperscript{1}, Muyi Sun\textsuperscript{2,$\dagger$}\\
\textsuperscript{1}Beijing University of Posts and Telecommunications, Beijing, China\\
\textsuperscript{2}Institute of Automation, Chinese Academy of Sciences, Beijing, China\\
\textsuperscript{3}Henan University of Chinese Medicine, Henan, China\\
{\tt\small \{zhuojiewu, wangzijianbupt, zouwenxuan, wanting.zhou\}@bupt.edu.cn} \\
{\tt\small \{fan.ji, muyi.sun\}@cripac.ia.ac.cn, danglee@hactcm.edu.cn} \\
}

\maketitle

\begin{abstract}
3D to 2D retinal vessel segmentation is a challenging problem in Optical Coherence Tomography Angiography (OCTA) images. Accurate retinal vessel segmentation is important for the diagnosis and prevention of ophthalmic diseases. However, making full use of the 3D data of OCTA volumes is a vital factor for obtaining satisfactory segmentation results. In this paper, we propose a Progressive Attention-Enhanced Network (PAENet) based on attention mechanisms to extract rich feature representation. Specifically, the framework consists of two main parts, the three-dimensional feature learning path and the two-dimensional segmentation path. In the three-dimensional feature learning path, we design a novel Adaptive Pooling Module (APM) and propose a new Quadruple Attention Module (QAM). The APM captures dependencies along the projection direction of volumes and learns a series of pooling coefficients for feature fusion, which efficiently reduces feature dimension. In addition, the QAM reweights the features by capturing four-group cross-dimension dependencies, which makes maximum use of 4D feature tensors. In the two-dimensional segmentation path, to acquire more detailed information, we propose a Feature Fusion Module (FFM) to inject 3D information into the 2D path. Meanwhile, we adopt the Polarized Self-Attention (PSA) block to model the semantic interdependencies in spatial and channel dimensions respectively. Experimentally, our extensive experiments on the OCTA-500 dataset show that our proposed algorithm achieves state-of-the-art performance compared with previous methods.
\end{abstract}


\section{Introduction}
Retinal vessel segmentation is of great significance in the diagnosis of various ophthalmic diseases, such as diabetic retinopathy and glaucoma, which can lead to blindness \cite{b1}. In clinical practice, accurate retinal vessel segmentation could help doctors accurately diagnose diseases and improve diagnostic efficiency.  
\begin{figure}[htbp]
\centerline{\includegraphics[scale=0.4]{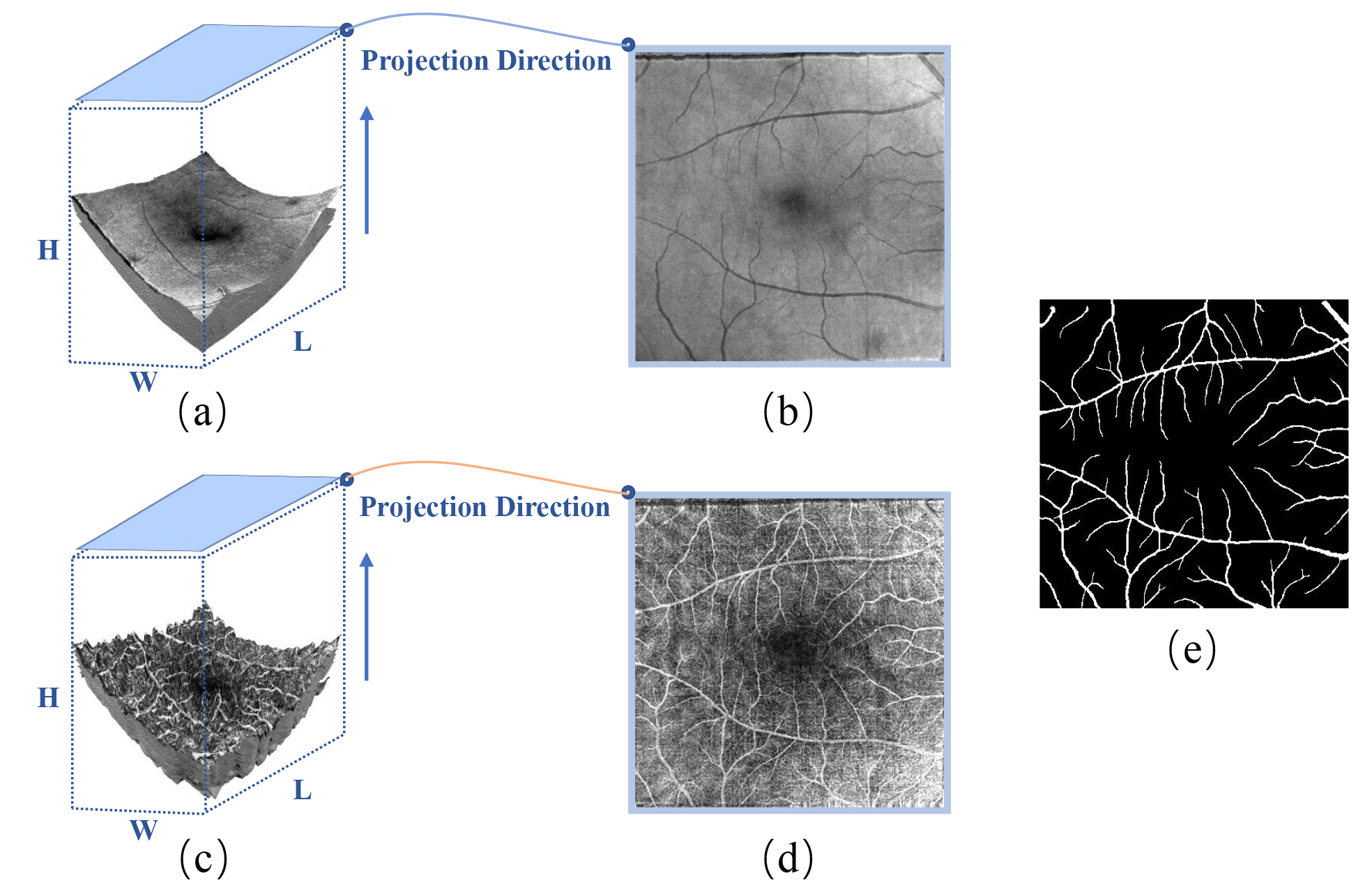}}
\caption{(a) OCT volume. (b) projection map of OCT. (c) OCTA volume. (d) projection map of OCTA. (e) ground truth.}
\label{fig1}
\end{figure}

Optical Coherence Tomography (OCT) is a novel and noninvasive optical imaging modality, which uses coherent light to capture 3D structural data of retina with micrometer-level resolution \cite{b2}, as shown in Fig.~\ref{fig1}(a). Compared with color fundus images, OCT volumes can provide detailed information about the structure of the retina. Meanwhile, OCTA volumes can provide rich information about retinal blood flow, which is generated from the OCT volumes by the Split-Spectrum Amplitude-Decorrelation Angiography (SSADA) algorithm \cite{b3}, as shown in Fig.~\ref{fig1}(c). OCTA solves the problem of OCT's inability to provide specific details about blood flow. Therefore, OCTA has become an important imaging method for the clinical diagnosis of retinal diseases and got a lot of attention from researchers nowadays.

Most of the previous work focused on color fundus images for retinal vessel segmentation. The traditional methods of retinal vessel segmentation are mostly accomplished by manually designed features \cite{b4,b5,b6}. Recently, with the development of deep learning, various Convolutional Neural Networks (CNNs) are used for retinal vessel segmentation \cite{b7,b8,b9,b10,b11,b12,b13}. Before Fully Convolutional Networks were widely used, segmentation was regarded as a pixel-by-pixel classification task using fully connected networks to predict the label of the center pixel of each patch \cite{b7}. Subsequently, a method based on fully convolutional neural networks solve the structured prediction problem and accomplish end-to-end segmentation \cite{b8,b9}. Further development, encoder-decoder architecture, especially U-Net architecture, has become the most popular segmentation framework for fundus images due to its excellent feature extraction capability and prominent practical performance  \cite{b10,b11,b12}. Various sophisticated models based on encoder-decoder architecture are designed to solve the problems of retinal blood vessel segmentation \cite{b10,b11,b12}. To improve performance on capillaries, the multi-label architecture is proposed by adding additional supervisions to treat thin and thick vessels respectively \cite{b10}. Meanwhile, a coarse to fine network consisting of two U-shaped networks is designed for vessel segmentation. The coarse network produces a preliminary prediction map and the fine network refines the results \cite{b11}. In addition, the cross-connected multi-scale network solves the segmentation problem of fine vessels \cite{b12}. 

In recent years, the attention mechanisms has also been widely used in retinal vessel segmentation because of its powerful feature-dependent modeling capabilities \cite{b13,b14,b15,b16}. RNA-Net employs residual non-local attention to solve the problem of the local fixed receptive fields which unable to collect global information \cite{b13}. In addition, a spatial attention lightweight network, named SA-UNet, enables efficient use of samples when only a few labeled samples are available \cite{b14}. Besides, CGA-Net designs a context guided attention for the imbalance of retinal vessel thickness distribution \cite{b15}. For the multi-scale vessel structure, Fully Attention-based Network (FANet) introduces the dual-direction attention into the retinal vessel segmentation \cite{b16}.

Following the above methods, retinal vessel segmentation of OCTA volumes firstly require the obtainment of the corresponding projection image (Fig.~\ref{fig1}(b) and Fig.~\ref{fig1}(d)) by retinal layer segmentation. However, some retinal diseases would destroy the retina structure and affect the retinal layer segmentation, in turn cause failed segmentation result. Moreover, segmentation on projection images cannot make full use of 3D information and bring OCTA volumes superiority into full play. After a long period of unremitting efforts, some progresses have been made in directly using OCTA volumes to obtain 2D vessel segmentation results. An Image Projection Network (IPN) achieves 3D to 2D image segmentation by employing unidirectional pooling along the projection direction of volumes \cite{b17}. In addition, Image Projection Network V2 (IPN-V2) and IPN-V2+ are proposed to enhance the ability of the horizontal direction perception and overcome the ``checkerboard effect" respectively \cite{b18}. The 3D to 2D retinal vessel segmentation methods provide a novel and promising research idea. However, the unidirectional pooling layer cannot capture 3D information well for image segmentation. Meanwhile, the plane network does not use volumetric data and lacks spatial-wise and channel-wise dependencies.

Inspired by the methods above mentioned, in this paper, we propose a Progressive Attention-Enhanced Network (PAENet) for 3D to 2D retinal vessel segmentation. The framework consists of two main parts, the three-dimensional feature learning path, and the two-dimensional segmentation path. To fuse 3D features more effectively, we design a novel Adaptive Pooling Module (APM), which learns a series of pooling coefficients along the projection direction of the volumes to fuse features adaptively. Due to the lack of volumetric data reuse in the two-dimensional segmentation path, we propose a Feature Fusion Module (FFM) to inject 3D information into the 2D path. Furthermore, in order to learn rich feature representation, in the three-dimensional feature learning path, we propose a new Quadruple Attention Module (QAM) that captures cross-dimension dependencies. In the two-dimensional segmentation path, we adopt the Polarized Self-Attention (PSA) block to model the semantic interdependencies in spatial and channel dimensions respectively.

The main contributions are summarized as follows:
\begin{itemize}
\item We propose a novel Progressive Attention-Enhanced Network (PAENet) to accomplish 3D to 2D segmentation and a Feature Fusion Module (FFM) which accomplishes volumetric data reuse.
\item Adaptive Pooling Module (APM) is proposed to capture dependencies along the projection direction of volumes and effectively reduce the feature dimension in an adaptive way.
\item We propose a new Quadruple Attention Module (QAM) to capture cross-dimension dependencies, and utilize Polarized Self-Attention (PSA) block to model the semantic interdependencies in spatial and channel dimensions respectively.
\item Sufficient experiments are conducted on the OCTA-500 dataset, the results demonstrate that our proposed method achieves state-of-the-art performance.
\end{itemize}
\begin{figure*}[t]
\centerline{\includegraphics[scale=0.5]{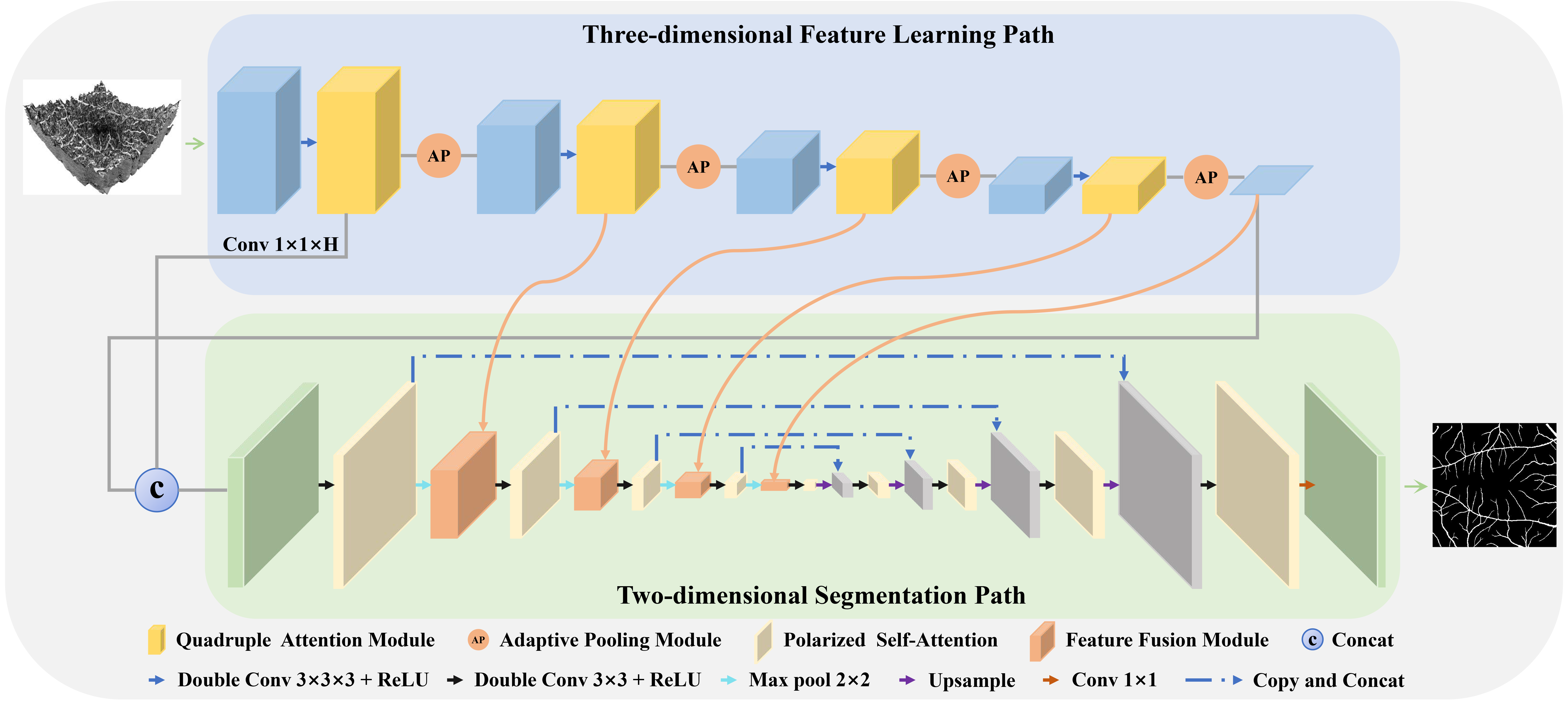}}
\caption{The framework of Progressive Attention-Enhanced Network (PAENet).}
\label{fig2}
\end{figure*}
\section{Related Work}
\subsection{Retinal Vessel Segmentation}
Retinal vessel segmentation can help doctors diagnose various ophthalmic diseases. Previous work mostly uses 2D color fundus images. Most of the early methods are manually design features for retinal vessel segmentation \cite{b4,b5,b6}. Staal \emph{et al.} \cite{b19} propose a method based on image ridge to solve the automatic blood vessel segmentation. However, the prior methods have relatively poor robustness. With the rapid development of deep learning in medical image segmentation, the methods based on deep learning have shown strong robustness and powerful feature extraction ability \cite{b16,b20,b21}. Fu \emph{et al.} \cite{b20} propose DeepVessel for accurate segmentation of capillaries and vessel junctions. This method uses a multi-scale and multi-level network to learn a rich hierarchical representation and employs a Conditional Random Field (CRF) to model the long-range interactions between pixels. Nevertheless, the networks' ability to recognize fine blood vessels is still limited. Guo \emph{et al.} \cite{b21} embeds channel attention into U-Net to enhance the discrimination ability of the network. However, the segmentation result is coarse. Li \emph{et al.} \cite{b16} propose the FANet based on attention mechanisms, which designed a dual-direction attention block to model global dependencies, making the segmentation result more delicate.

Compared with color fundus images, OCT volumes can provide detailed information about the structure of the retina. Researchers begin to study the use of OCT volumes for retinal vessel segmentation. Li \emph{et al.} \cite{b17} propose an image projection network that accomplishes 3D to 2D image segmentation by unidirectional pooling along the projection direction of volumes. In addition, Image Projection Network V2 \cite{b18} is proposed to enhance the ability of the horizontal direction perception. However, the unidirectional pooling layer cannot adaptively capture 3D information for image segmentation. Meanwhile, the plane perceptron does not use volumetric data and lacks spatial-wise and channel-wise dependencies. In this paper, we promote the idea of the above method and propose the APM for efficient fuse volumetric data and reduce the feature dimension. The FFM is also proposed for volumetric data reuse in the two-dimensional segmentation path.
\subsection{Attention Mechanism}
Attention mechanism can effectively help the model refining perceive information and have proven to be helpful in computer vision tasks \cite{b22,b23,b24}. Squeeze-and-Excitation Networks (SENet) \cite{b22} enhances the representational ability of CNN by modeling the correlation between feature channels, while SENet lacks spatial attention. Woo \emph{et al.} \cite{b23} propose a Convolutional Block Attention Module (CBAM) which cascade channel-wise attention and spatial-wise attention. However, the dependency of CBAM is local. Wang \emph{et al.} \cite{b24} propose a Non-local Neural Network which computes the response at a position as a weighted sum of the features at all positions, so as to model global dependence. Fu \emph{et al.} \cite{b25} introduces DANet which captures the semantic interdependencies in spatial and channel dimensions respectively. However, the computational cost of DANet is very huge because it computes the relationship between each location and each channel. Misra \emph{et al.} \cite{b26} propose a lightweight but effective triplet attention network that captures cross-dimensional dependencies through three branches. However, the triple attention network can only be used in 2D networks, and can not effectively capture the cross-dimensional dependence of 3D networks. In this paper, we propose a Quadruple Attention Module (QAM) for capture cross-dimension dependencies of 4D feature tensors. Meanwhile, we apply Polarized Self-Attention (PSA) block \cite{b27} to effectively model global dependence of space-wise and channel-wise.
\section{Method}
In this work, we develop a Progressive Attention-Enhanced Network (PAENet) for 3D to 2D retinal vessel segmentation. The framework consists of two main parts, the three-dimensional feature learning path and the two-dimensional segmentation path. Specifically, the three-dimensional feature learning path mainly uses APM for 3D volume dimension reduction and QAM for capturing cross-dimensional feature representation. In addition, the two-dimensional segmentation path is a progressive U-Net integrated with the PSA block and injected with 3D information. The framework of PAENet is shown in Fig.~\ref{fig2}. Since we input OCT volume and OCTA volume together as two channels, assuming that the volume size is $\left ( {L\times W\times H} \right )$, and then the input of PAENet is $X\in \mathbb{R}^{C\times L\times W\times H}$ and the output is $Y\in \mathbb{R}^{L\times W}$. Here C equals 2.
\subsection{Adaptive Pooling Module}
\begin{figure*}[h]
\centerline{\includegraphics[scale=0.6]{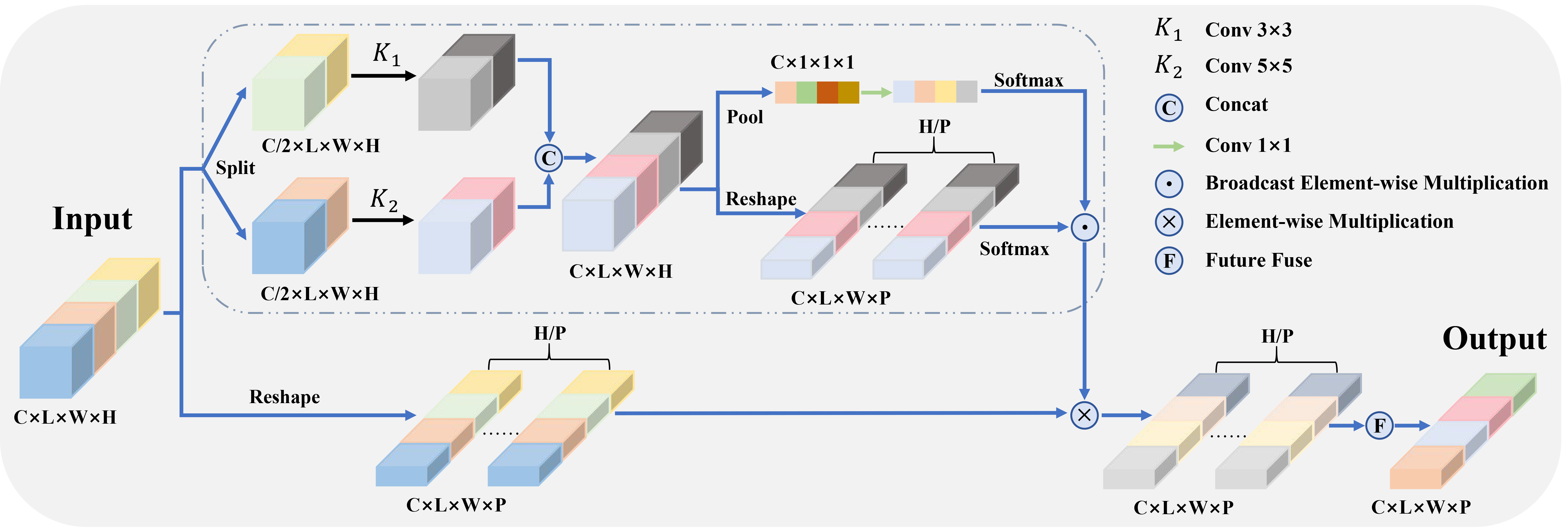}}
\caption{The structure of the Adaptive Pooling Module.}
\label{fig3}
\end{figure*}
In the three-dimensional feature learning path, it is very important to effectively reduce feature dimension and retain information to the greatest extent. Therefore, we propose Adaptive Pooling Module (APM) to achieve the above purpose. The structure of APM is shown in Fig.~\ref{fig3}.

The input and output of APM are $I\in \mathbb{R}^{C\times L\times W\times H}$ and $y\in \mathbb{R}^{C\times L\times W\times P}$ respectively. P is the pooling size. To begin with, we divide the input into two groups and extract multi-scale features through two branches. The feature of each branch is $I_{i}\in \mathbb{R}^{\frac{C}{2}\times L\times W\times H},i=1,2$. $K1$ and $K2$ are two filters with kernel sizes of 3×3 and 5×5 respectively. Next, the whole multi-scale feature map $M$ can be obtained by concatenation. The process is shown in the following formula:
\begin{equation}
M= Concat(K_{1}(I_{1}),K_{2}(I_{2})) \label{eq1}
\end{equation}
where $M\in \mathbb{R}^{C\times L\times W\times H}$.

In the APM, we not only focus on spatial information, but also capture the correlation of channels. The channel description is obtained from the spatial information of multi-scale feature $M$ using global average pooling. Next, the channel-wise dependencies are captured by the convolutional operation. The channel attention $Z$ is defined as:
\begin{equation}
Z=\sigma (Conv_{2}(\delta (Conv_{1}(AvgPool(M))))) \label{eq2}
\end{equation}
where $Z\in \mathbb{R}^{ C\times 1\times 1\times 1}$, $\sigma$ denotes the sigmoid function, $\delta$ represents the ReLU function, $Conv_{1}$ and $Conv_{2}$ is the 1×1 convolution.

To facilitate the subsequent feature fusion, the multi-scale feature map $M$ is reshaped to ${M}'\in \mathbb{R}^{\frac{H}{P}\times C\times L\times W\times P}$. In addition, the same operation is performed on input $I$ to obtain ${I}'\in \mathbb{R}^{\frac{H}{P}\times C\times L\times W\times P}$. After reweighting ${I}'$, summation is conducted along the projection direction of the features for feature fusion. The final output of APM can be written as follows:
\begin{equation}
y=F({I}'\cdot (Softmax ({M}')\bigodot Softmax (Z))) \label{eq3}
\end{equation}
where the $Softmax$ is used to obtain the attention weights in projection direction and channel dimension. $\bigodot$ denotes broadcast element-wise multiplication, and $\cdot$ refers to element-wise multiplication. $F$ refers to the feature fusion along the projection direction.
\subsection{Quadruple Attention Module}
\begin{figure}[h]
\centerline{\includegraphics[scale=1]{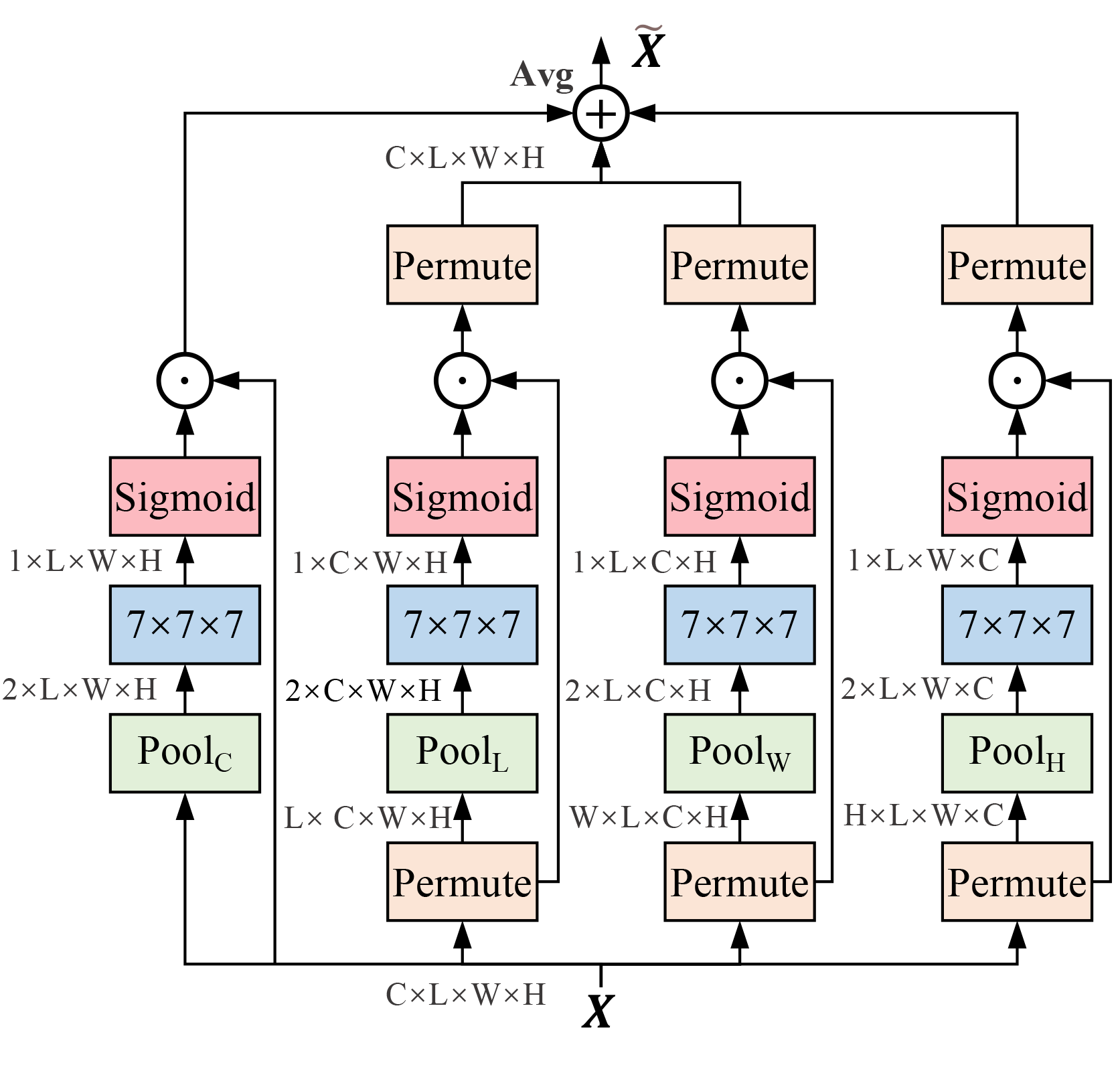}}
\caption{Schematic illustration of the proposed Quadruple Attention Module. The subscript of the pool indicates that the pooling operation is performed in the corresponding first dimension. $\bigodot$ denotes broadcast element-wise multiplication, and $\bigoplus$ denotes average broadcast element-wise addition.}
\label{fig4}
\end{figure}
To capture cross-dimensional dependencies on the 4D tensor, we propose a lightweight and effective Quadruple Attention Module (QAM). The QAM does not change the dimension size of the input. Suppose the input is $X\in \mathbb{R}^{C\times L\times W\times H}$ and the output is $\widetilde{X} \in \mathbb{R}^{C\times L\times W\times H}$, as shown in Fig.~\ref{fig4}. To begin with, the input $X$ is divided into four branches by permuting. Next, the pooling layer preserves the rich representation of features through the max pool and average pool and reduces the first dimension of input to two. For example, the input $X\in \mathbb{R}^{C\times L\times W\times H}$ permute to $\mathbb{R}^{L\times C\times W\times H}$, and then the final result $\mathbb{R}^{2\times C\times W\times H}$ is obtained by concatenating the max pooled features and the average pooled features on the first dimension. This process can be represented by the following equation:
\begin{equation}
Pool_{d}({X}')=Concat(MaxPool_{d}({X}'),AvgPool_{d}({X}')) \label{eq4}
\end{equation}
where ${X}'$ refers to the permuted input. $d$ represents the first dimension of permuted tensor. Behind the pooling layer is a standard convolutional layer with the convolution kernel size $7\times 7\times 7$ followed by a batch normalization layer and ReLU. An attention weight map is obtained through the sigmoid activation function.

Concretely, in the first branch, the relationship between the three dimensions (L, W, H) is constructed. The attention weight map ${\widetilde{X}}'_{1} \in \mathbb{R}^{1\times L\times W\times H}$ obtained through the pooling layer and convolution layer is multiplied by the input ${X}'_{1} \in \mathbb{R}^{C\times L\times W\times H}$. The output of this branch is $y_{1} \in \mathbb{R}^{C\times L\times W\times H}$. Meanwhile, the second branch constructs the relationship of three dimensions (C, W, H) by permuted. The tensor ${X}'_{2} \in \mathbb{R}^{L\times C\times W\times H}$ after permuted obtains the attention weight map ${\widetilde{X}}'_{2} \in \mathbb{R}^{1\times C\times W\times H}$ through the pooling layer and convolutional layer. Then the attention weight map is multiplied by the input to obtain the final output $y_{2} \in \mathbb{R}^{L\times C\times W\times H}$. For the remaining branches, the relationship of (L, C, H) and (L, W, C) dimensions is constructed respectively. It is worth noting that before adding the results of the last three branches to the results of the first branch, they should be permuted and restored to the original input shape. Therefore, the final output of QAM is mathematically expressed as: 
\begin{equation}
\widetilde{X}=\frac{1}{4}(y_{1}+\bar{y_{2}}+\bar{y_{3}}+\bar{y_{4}}) \label{eq5}
\end{equation}
where $\bar{}$ means the permute operation.
\subsection{Feature Fusion Module}
\begin{figure}[h]
\centerline{\includegraphics[scale=1]{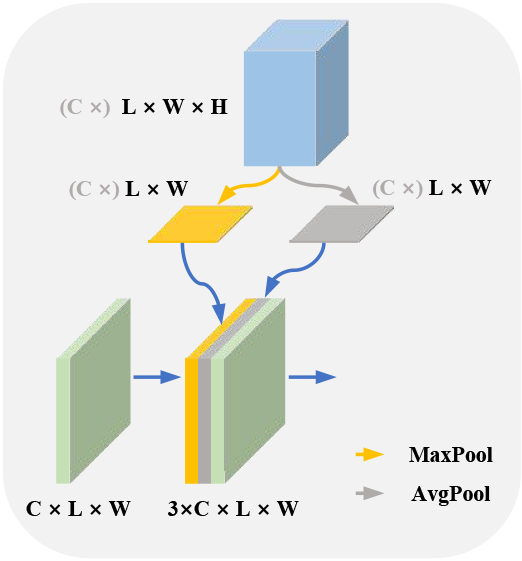}}
\caption{The illustration of Feature Fusion Module. Since the 4D tensor is not shown in the figure, the channel dimension is represented in gray font.}
\label{fig5}
\end{figure}
To utilize the information of the 3D feature learning path in the 2D segmentation path, the Feature Fusion Module (FFM) is proposed to inject the feature maps of the 3D feature learning path into the 2D segmentation path for improving the performance of network and accomplishing volumetric data reuse. Concretely, as shown in Fig.~\ref{fig5}, the 4D tensor $X \in \mathbb{R}^{C\times L\times W\times H}$ is compressed to the same shape as 3D tensor $\widetilde{X} \in \mathbb{R}^{C\times L\times W}$ by pooling layers. Next, the pooled features are connected with the features of the 2D segmentation path. The MaxPool preserves more texture features, and the AvgPool preserves more overall features. Mathematically, it can be represented by the following equation:
\begin{equation}
f(X,\widetilde{X})=Concat(\widetilde{X},AvgPool(X),MaxPool(X)) \label{eq6}
\end{equation}
\subsection{Lightweight Polarized Self-Attention}
\begin{figure}[h]
\centerline{\includegraphics[scale=1.1]{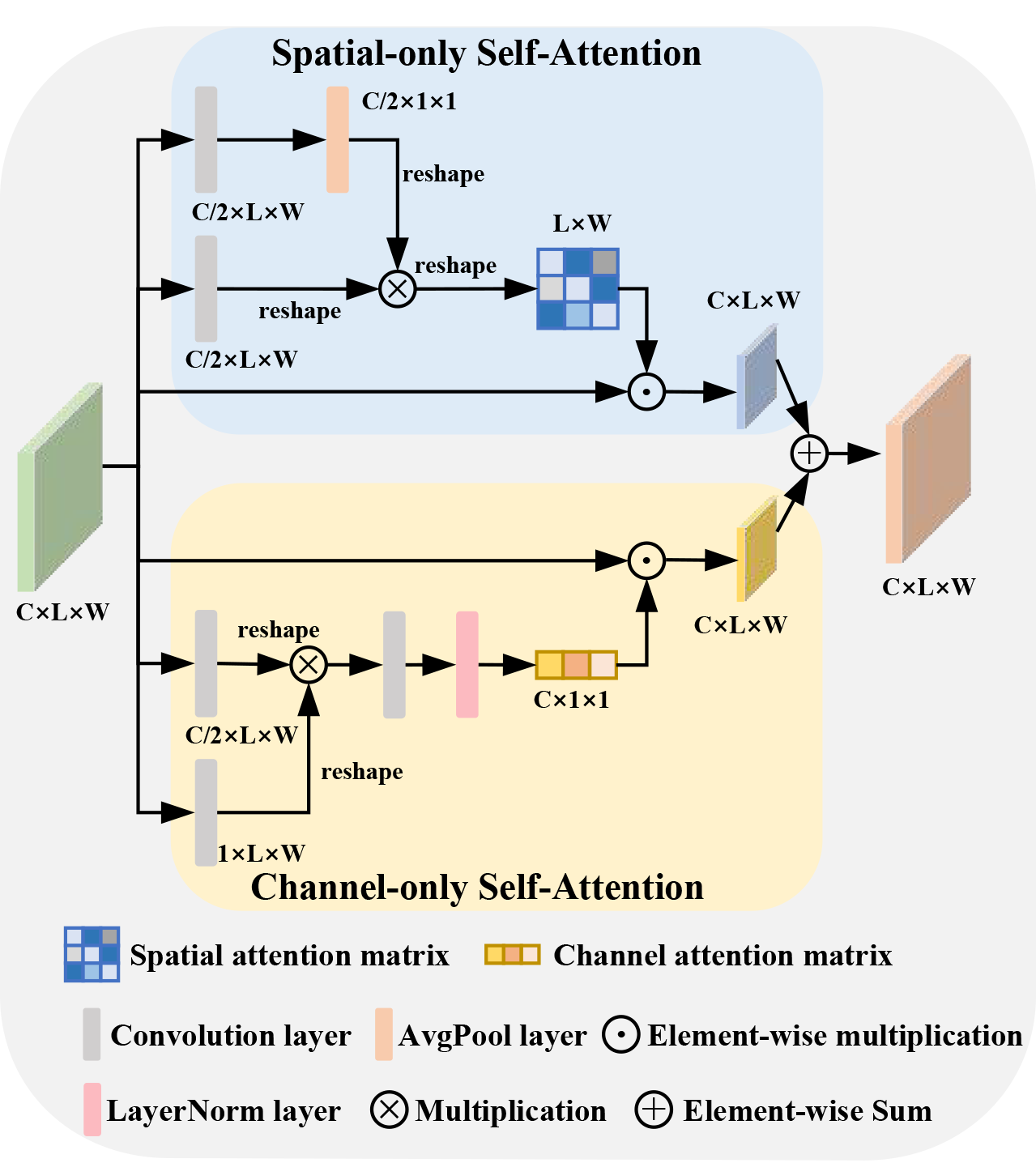}}
\caption{The illustration of the Polarized Self-Attention block.}
\label{fig6}
\end{figure}
In the 2D segmentation path, we employ the lightweight Polarized Self-Attention (PSA) block that capture long-range contextual information in space and channel dimensions to boost the performance of the plain U-net. The PSA block is shown in Fig.~\ref{fig6}. The PSA block has two branches: Spatial-only Self-Attention and Channel-only Self-Attention.
\subsubsection{Spatial-only Self-Attention}
The branch weights the input by generating a spatial attention matrix. Mathematically, it can be expressed as the following formula:
\begin{equation}
f_{sp}(X)=\Gamma (Softmax(\Gamma (Avg(W_{1}X))\times \Gamma (W_{2}X))\cdot X \label{eq7}
\end{equation}
where $W_{1}$ and $W_{2}$ are $1\times 1$ convolutional layers respectively. $\Gamma$ is reshape operator. $\times$ and $\cdot$ represent matrix multiplication and broadcast element-wise multiplication, respectively.
\subsubsection{Channel-only Self-Attention}
The weighting channel-wise is shown in the following formula: 
\begin{equation}
f_{ch}(X)=LN(W_{3}(Softmax(\Gamma W_{1}X)\times \Gamma(W_{2}X)))\cdot X \label{eq8}
\end{equation}
where $W_{1}$, $W_{2}$ and $W_{3}$ are $1\times 1$ convolutional layers respectively. $LN$ is LayerNorm layer.
Therefore, the final output of the PSA block is as follows:
\begin{equation}
PSA(X)=f_{sp}(X)+f_{ch}(X) \label{eq9}
\end{equation}
where $+$ is element-wise sum.
\begin{table*}[h]
\centering
\caption{Effect of different modules in PAENet (MEAN ± SD). APM: Adaptive Pooling Module. QAM: Quadruple Attention Module. FFM: Feature Fusion Module. PSA: Polarized Self-Attention.}
\begin{tabular}{lllllllcc}
\hline
\multicolumn{1}{c}{APM} & QAM          & FFM          & PSA                               & \multicolumn{1}{c}{DICE(\%)} & \multicolumn{1}{c}{JAC(\%)} & \multicolumn{1}{c}{BACC(\%)} & PRE(\%)                          & REC(\%)                                   \\ \hline
                        &              &              & \multicolumn{1}{l|}{}             & 89.10 ± 2.82                 & 80.45 ± 4.36                & 93.85 ± 1.87                 & 89.57 ± 3.69                     & 88.77 ± 3.69                              \\
$\checkmark$            &              &              & \multicolumn{1}{l|}{}             & 89.23 ± 2.79                 & 80.67 ± 4.37                & 93.83 ± 2.09                 & \textbf{89.95 ± 3.22}            & 88.67 ± 4.19                              \\
$\checkmark$            & $\checkmark$ &              & \multicolumn{1}{l|}{}             & 89.27 ± 2.82                 & 80.73 ± 4.40                & 93.91 ± 1.98                 & 89.83 ± 3.56                     & 88.87 ± 3.97                              \\
$\checkmark$            & $\checkmark$ & $\checkmark$ & \multicolumn{1}{l|}{}             & 89.31 ± 2.58                 & 80.79 ± 4.09                & 93.98 ± 1.98                 & \multicolumn{1}{l}{89.77 ± 2.96} & \multicolumn{1}{l}{89.00 ± 4.02}          \\
$\checkmark$            & $\checkmark$ & $\checkmark$ & \multicolumn{1}{l|}{$\checkmark$} & \textbf{89.36 ± 2.70}        & \textbf{80.87 ± 4.25}       & \textbf{94.04 ± 1.95}        & \multicolumn{1}{l}{89.73 ± 3.32} & \multicolumn{1}{l}{\textbf{89.14 ± 3.91}} \\ \hline
\end{tabular}
\label{tab1} 
\end{table*}

\begin{table*}[h]
\centering
\caption{comparison results with existing methods (MEAN ± SD). No. 1 denotes that the global retraining process is not introduced. No. 2 denotes that the global retraining process is introduced}
\begin{tabular}{c|l|lllcc}
\hline
No.                & \multicolumn{1}{c|}{Methods} & \multicolumn{1}{c}{DICE(\%)} & \multicolumn{1}{c}{JAC(\%)} & \multicolumn{1}{c}{BACC(\%)} & PRE(\%)                                   & REC(\%)                                   \\ \hline
\multirow{3}{*}{1} & IPN {[}17{]}                 & 88.64 ± 3.21                 & 79.73 ± 4.92                & 93.07 ± 2.42                 & -                                         & -                                         \\
                   & IPN V2 {[}18{]}              & 89.08 ± 2.73                 & 80.41 ± 4.29                & 93.52 ± 2.13                 & -                                         & -                                         \\
                   & PAENet                       & \textbf{89.36 ± 2.70}        & \textbf{80.87 ± 4.25}       & \textbf{94.04 ± 1.95}        & \multicolumn{1}{l}{\textbf{89.73 ± 3.32}} & \multicolumn{1}{l}{\textbf{89.14 ± 3.91}} \\ \hline
\multirow{2}{*}{2} & IPN V2+ {[}18{]}             & 89.41 ± 2.74                 & 80.95 ± 4.32                & 93.46 ± 2.12                 & -                                         & -                                         \\
                   & PAENet+                      & \textbf{89.69 ± 2.77}        & \textbf{81.42 ± 4.39}       & \textbf{93.68 ± 2.08}        & \multicolumn{1}{l}{\textbf{91.37 ± 3.23}} & \multicolumn{1}{l}{\textbf{88.22 ± 4.19}} \\ \hline
\end{tabular}
\label{tab2} 
\end{table*}
\section{Experiments}
\subsection{Datasets}
We select data with a field of view size of 6mm $\times$ 6mm on the public dataset OCTA-500 to evaluate PAENet \cite{b18}. The dataset contains 300 subjects (NO.10001-NO.10300) with the volume size of 400px × 400px × 640px. Each sample contains a pair of volume data of OCT and OCTA. Following the previous work \cite{b18}, the dataset is divided into a training set (NO.10001- NO.10180), a validation set (NO.10181-NO.10200), and a test set (NO.10201-NO.10300).
\subsection{Implementation Details}
The proposed network is implemented on the pytorch platform with two TITAN Xp GPUs. We employ Adaptive Moment Estimation (Adam) optimization with momentum 0.9. Meanwhile, we utilize the poly learning rate policy \cite{b28,b29}. The initial learning rate is set to 0.0003, and the batchsize is set to 4. Additionally, We crop each volume into patches for training. The patch size is 100px × 100px × 160px, the total training iterations are 25000.
\subsection{Evaluation Metrics}
To assess the performance of our proposed network, we choose five metrics for evaluation: dice coefficient (DICE), jaccard index (JAC), balance-accuracy (BACC), precision (PRE) and recall (REC).
\begin{equation}
DICE=2TP / (2TP + FP + FN) \label{eq10}
\end{equation}
\begin{equation}
JAC=TP/(TP + FP + FN) \label{eq11}
\end{equation}
\begin{equation}
BACC=(TPR+TNR)/2 \label{eq12}
\end{equation}
\begin{equation}
PRE=TP/(TP+FP) \label{eq13}
\end{equation}
\begin{equation}
REC=TP/(TP+FN) \label{eq14}
\end{equation}
where $TP$ is true positive, $FP$ is false positive, $TN$ is true negative and $FN$ is false negative. As we all know, the proportion of foreground and background is seriously unbalanced in retinal vessel segmentation. It is not reasonable to use accuracy to evaluate the segmentation results. Therefore, we use balance-accuracy to evaluate the results for the imbalance of the positive and negative samples. In (\ref{eq12}), $TPR=TP/(TP+FN)$ is true positive rate, and $TNR=TN/(TN+FP)$ is true negative rate. The experimental result is the average and standard deviation on the test set.
\subsection{Ablation Study}
\begin{figure*}[h]
\centerline{\includegraphics[scale=1]{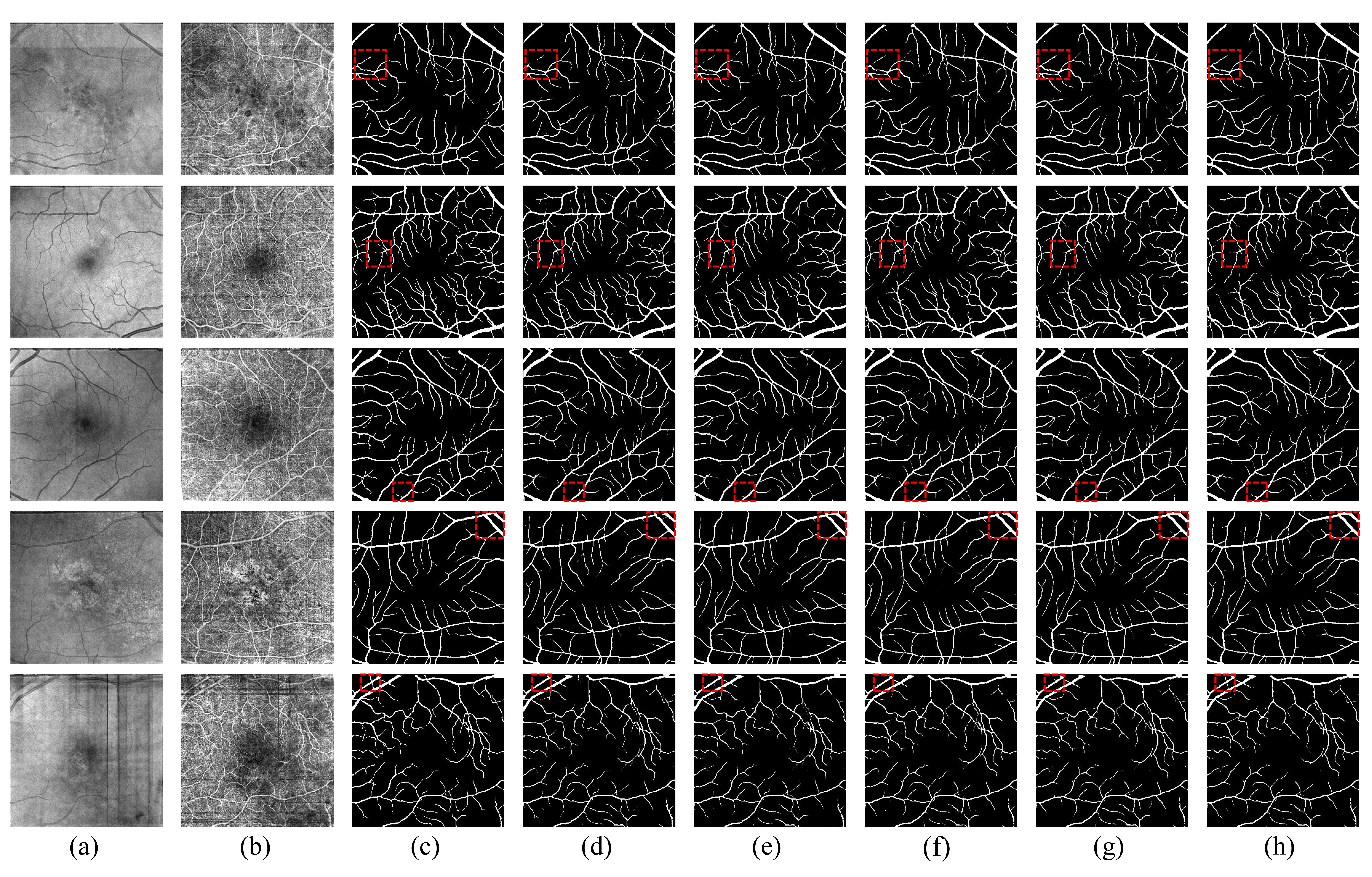}}
\caption{Qualitative results of our proposed method. (a) The projection map of OCT, (b) The projection map of OCTA, (c) The corresponding ground truth, (d) The result of baseline, (e) The result of adding APM, (f) The result of adding APM and QAM, (g) The result of adding APM, QAM and FFM, (h) The results of the proposed PAENet. The red box is the details of the segmentation. From (d) to (h), the segmentation results are getting better and better.}
\label{fig7}
\end{figure*}
\subsubsection{Ablation Study on Module}
We conduct a large number of experiments to verify the effectiveness of our proposed modules. We set up five groups of ablation experiments, and the experimental designs are listed as follows: (1) Baseline uses a unidirectional pooling \cite{b18} to reduce the feature dimension, while the next experiment uses APM instead. (2) only APM is integrated into baseline architecture. (3) APM and QAM are jointly integrated into baseline architecture. (4) APM, QAM and FFM are jointly integrated into baseline architecture. (5) All modules are jointly integrated into baseline architecture. The experimental results are shown in Table.~\ref{tab1}. In the second set of experiments, we can observe that adaptive pooling improves the baseline by 0.13$\%$ (DICE) / 0.22$\%$ (JAC) / 0.38$\%$ (PRE) and shows the highest PRE results. It is proved that the APM can effectively fuse volume information. Compared with unidirectional pooling, the APM has a stronger feature fusion ability that enables the network to recognize unlabeled capillaries, which leads to a slight decrease in BACC and REC. However, the problem mentioned above will be alleviated through model cross-dimensional dependencies in the subsequent modules. When APM and QAM are added to the baseline architecture, the improvements are 0.17$\%$ (DICE) / 0.28$\%$ (JAC) / 0.06$\%$ (BACC) / 0.26$\%$ (PRE) / 0.10$\%$ (REC). Obviously, the effectiveness of QAM in capturing cross-dimensional dependencies is proved. In the fourth group of experiments with the APM, QAM and FFM, the FFM injects volume information into the 2D segmented network, which improves the network by 0.21$\%$ (DICE) / 0.34$\%$ (JAC) / 0.13$\%$ (BACC) / 0.20$\%$ (PRE) / 0.23$\%$ (REC). According to the above data, it is proved that volumetric data reuse can provide more detailed information to the network. In the last group of experiments with all modules, the improvements are 0.26$\%$ (DICE) / 0.42$\%$ (JAC) / 0.19$\%$ (BACC) / 0.16$\%$ (PRE) / 0.37$\%$ (REC). Therefore, the PSA block is proved to be effective through modeling global spatial and channel dependence. Meanwhile, table.~\ref{tab1} shows that each module can coordinate with the other. The qualitative results of module ablation are shown in Fig.~\ref{fig7}. In the qualitative results of the ablation experiment, we can see that the results of the retina are getting better and the segmentation of fine blood vessels achieves better performance.
\subsubsection{Comparison with the Existing Methods}
The comparison results between our proposed method with existing methods are shown in Table.~\ref{tab2}. Obviously, our proposed PAENet achieves state-of-the-art performance compared with several existing methods. Compared with IPN, our methods outperforms the performance of the IPN by 0.72$\%$ (DICE) / 1.14$\%$ (JAC) / 0.97$\%$ (BACC). Our method also surpasses IPN V2 by 0.28$\%$ (DICE) / 0.46$\%$ (JAC) / 0.52$\%$ (BACC). It demonstrates that our proposed method can complete the 3D to 2D retinal vessel segmentation in OCTA images better. Following the previous work \cite{b18}, we also introduced a global retraining process and conducted related experiments. The experimental results are shown in Table.~\ref{tab2}-2. Compared with IPN v2+, our method outperforms the IPN v2+ by 0.28$\%$ (DICE) / 0.47$\%$ (JAC) / 0.22$\%$ (BACC) and reaches 89.69$\%$ (DICE) / 81.42$\%$ (JAC) / 93.68$\%$ (BACC). The above experiments show that our method can make full use of 3D volume data and model the dependence of different dimensions to accomplish the task of 3D to 2D retinal segmentation.
\subsection{Results}
We perform ablation analysis on each module to prove the effectiveness of the module, and the results are shown in Table.~\ref{tab1}. Meanwhile, the qualitative results of the module are shown in Fig.~\ref{fig7}. In Table.~\ref{tab2}, our method outperforms the performance of the existing methods. Extensive experimental results demonstrate that our method can effectively accomplish the task of 3D to 2D retinal segmentation in OCTA images and achieve state-of-the-art performance compared with several existing methods. In addition, we also conducted a qualitative analysis, as shown in Fig.~\ref{fig7}. Our method not only performs accurate segmentation of some fine vessels, but also completes the segmentation efficiently when the input image quality is poor.
\section{Conclusion}
In this paper, we propose a novel Progressive Attention-Enhanced Network (PAENet) for 3D to 2D retinal vessel segmentation in OCTA images. Specifically, we propose a novel Adaptive Pooling Module which captures dependencies along the projection direction of volumes for feature fusion. Meanwhile, we design a Quadruple Attention Module to model the cross-dimension relationship in the 4D tensor. To make full use of volume information, FFM is introduced to inject 3D information into the 2D segmentation network to accomplish volumetric data reuse. In addition, Polarized Self-attention blocks are integrated into the network to model global spatial and channel attention. Extensive experiments demonstrate that PAENet is an effective implementation of 3D to 2D segmentation networks and achieves state-of-the-art performance compared with previous methods. In the future, we will explore the PAENet structure to achieve better performance and solve the problems of artifacts and capillaries in OCTA images.
\section*{Acknowledgment}
This work is completed when Zhuojie Wu is an intern under the guidance of Dr. Muyi Sun. This work is supported by NSFC (Grant No. 62006227).

\end{document}